\documentclass[prl,aps,epsfig,twocolumn,showpacs,amsmath,amssymb]{revtex4} 
 
\usepackage{graphicx} 
\usepackage[dvips]{epsfig}
\usepackage{float}

\newcommand{\be}{\begin{equation}}
\newcommand{\ee}{\end{equation}}

\newcommand{\Hb}{\protect{{\bf H}}}
\newcommand{\Eb}{\protect{{\bf E}}}

\newcommand{\Pb}{\protect{{\bf P}}}
\newcommand{\Mb}{\protect{{\bf M}}}

\newcommand{\Bb}{\protect{{\bf B}}}

\newcommand{\kb}{\protect{{\bf k}}}

\begin{document} 

\title{Comment on: "Electromagnetically induced left-handedness in optically excited four-level atomic media"
by: Quentin Thommen and Paul Mandel}
\author{J\"urgen K\"astel}
\affiliation{Fachbereich Physik, Technische Universit\"{a}t 
Kaiserslautern, D-67663 Kaiserslautern, 
Germany} 
\author{Michael Fleischhauer}
\affiliation{Fachbereich Physik, Technische Universit\"{a}t 
Kaiserslautern, D-67663 Kaiserslautern, 
Germany}

\date{\today} 
\begin{abstract}
\end{abstract}
\maketitle 

In a recent paper (Phys. Rev. Lett. {\bf 96}, 053601 (2006)) Thommen and Mandel discussed a novel scheme to induce
left-handedness and negative refraction in an atomic four-level scheme
\cite{Thommen06}. The proposal
is based on a coherent cross-coupling between electric and magnetic dipole transitions, which couple to the
electric and magnetic field components of the probe field. A very important feature of the Thommen-Mandel
scheme is that the two transitions do not have to involve common states, which greatly enhances the freedom
of choice of levels and makes the scheme much more applicable to realistic systems than previous
proposals \cite{Oktel}.  We here  show that although the main conclusion of \cite{Thommen06} -- the possibility to create
negative refraction in the coherently driven four-level scheme -- remains valid, the results obtained
are quantitatively not correct. 

 Let us consider the four-level scheme of ref.\cite{Thommen06} shown in 
Fig.1.
\begin{figure}[ht]
  \begin{center} 
    \epsfig{width=8cm,file=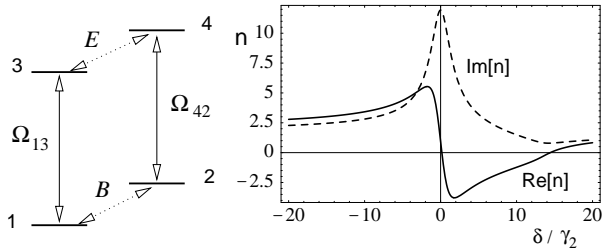} 
    \caption{{\it left:} 4-level scheme of \cite{Thommen06}, $\Omega_{13}$ and $\Omega_{42}$
denote the Rabi-frequencies of (strong) external drive fields. Electric ($E$) and magnetic ($B$)
components of the probe field couple to $|3\rangle-|4\rangle$ and
$|1\rangle-|2\rangle$ respectively. {\it right:} 
 Real (full line) and imaginary part (dashed line) of $n$. 
  The upper levels $|3\rangle$ and $|4\rangle$ of electric dipole transitions are assumed to decay with
  the rate $\gamma=$10$^7$ s$^{-1}$, while the upper level $|2\rangle$ of the magnetic transition decays
  with rate $\gamma_2=\gamma/(137)^2$. $\lambda=600$ nm, $\Omega_{13}=\gamma, \Omega_{42}=10^{-2}\gamma$.}
  \end{center} 
\label{fig1}
\end{figure} 
In order to obtain the results for the linear response, given in eqs.(4) and (5) of 
\cite{Thommen06}, Thommen and Mandel solved the density-matrix equations in 
the weak-excitation limit, i.e. setting  $\rho_{11}=1$. 
%
%
%
%
The imaginary part of the permittivity $\varepsilon$ obtained from this 
attains negative values
for certain parameter values, corresponding to gain, which is 
unphysical. 
This is a result of the weak-excitation assumption, which
is consistent only if the coupling to the external drive field 
$\Omega_{13}$ 
can be treated in first oder perturbation.
Solving the steady-state density matrix equations without this
approximation
leads to an imaginary part of $\varepsilon$ which is always positive. 

Furthermore the coherent cross-coupling between electric and magnetic dipole transitions
leads to chirality, where the magnetic 
component $\Hb$ of the probe field couples to the electric polarization $\Pb$ 
and correspondingly the electric component $\Eb$ to the
magnetization $\Mb$:
${\mathbf P} = \chi_{e}{\mathbf E}+\xi_{EH}{\mathbf H}$, and 
${\mathbf M}= \xi_{HE}{\mathbf E}+\chi_{m}{\mathbf H}$.
Here $\xi_{EH}$ and $\xi_{HE}$ are the chirality coefficients. 
In a chiral medium the expression for the index of refraction reads \cite{Pendry04}
\begin{equation}
n = \sqrt{\varepsilon\mu
-\frac{(\xi_{EH}+\xi_{HE})^2}{4}}
+\frac{i}{2}(\xi_{EH}-\xi_{HE}).\label{n-chiral}
\end{equation}
Rather than applying eq.(\ref{n-chiral}),
Thommen and Mandel followed the approach of Oktel and M\"ustecaplioglu \cite{Oktel}
 and used the relation $\Bb=\kb\times\Eb/(\omega c)$
to calculate the permeability $\mu$ from the matrix element $\rho_{12}(\Eb)$ and from that the
refractive index $n=\sqrt{\varepsilon\mu}$.  Although this captures the most important
contributions it neglects the modification of the electric polarization
by the magnetic component of the probe field.

We have calculated the index of refraction for the scheme of Fig.1 from eq.(\ref{n-chiral})  using the stationary
solutions of the density matrix equations without further assumptions. The results are shown on the right half of Fig.1,
where we have plotted Re$[n]$ and Im$[n]$ as function of the detuning $\delta=\omega-\omega_{34}
=\omega-\omega_{12}$, with $\omega$ being the probe-field frequency and $\omega_{12}=\omega_{34}$
the atomic transition frequencies.
We find that the imaginary part of the refractive index Im$[n]$ is always positive, corresponding to
absorption, and that the ratio of refraction to absorption, $\bigl|$Re$[n]/$Im$[n]\bigr|$,
is only on the order of unity.

\begin{acknowledgments}
J.K. acknowledges financial support by the Deutsche Forschungsgemeinschaft
through the 
GRK 792 ``Nichtlineare Optik und Ultrakurzzeitphysik''.
\end{acknowledgments}

\def\etal{\textit{et al.}}

\end{document}